\documentclass[12pt]{article}
\begin{document}

\title{Boosted Statistical Mechanics}
\author{Massimo Testa\\
Dipartimento di Fisica, Universit\`a degli Studi di Roma ``La Sapienza'' \\
and\\
INFN -- Sezione di Roma I, \\
Piazzale A. Moro 2, I-00185 Roma, Italy\\
massimo.testa@roma1.infn.it}
\date{}
\maketitle

\begin{abstract}
Based on the fundamental principles of Relativistic Quantum Mechanics, we give a rigorous, but completely elementary, proof of the relation between fundamental observables of a statistical system when measured relatively to two inertial reference frames, connected by a Lorentz transformation.
\end{abstract}

\section{Introduction}

The problem of the relation between the basic observables of a statistical mechanical system as seen from two inertial observers ${\cal O}$ and ${\cal O}'$, connected by a Lorentz transformation $x' = \Lambda x$, is very old and its solution can be found in classical textbooks\cite{pauli}. The detailed discussion and application of the basic principles, in the case of the black body radiation, followed the discovery of the cosmic microwave background\cite{heer}.

Although completely solved, the problem still stimulates scientific production\cite{ford}. It is the purpose of the present paper to show how the relation between inertial observers can be obtained from the basic principles of Relativisic Quantum Mechanics. We will also discuss in detail the transformation properties of quantities related to the polarization of observed photons.

\section{Boosted Density Matrix}

In Relativistic Quantum Theory\cite{bjorken} the relation between two inertial observers ${\cal O}$ and ${\cal O}'$, connected by a Lorentz transformation $x' = \Lambda x$, is provided by a unitary operator, $U(\Lambda)$, in the sense that, if $| \Psi >$ is the vector which describes the physical situation as seen from ${\cal O}$, the state $| \Psi' >$ attributed to the same physical configuration by ${\cal O}'$ is 
\begin{eqnarray}
| \Psi' > = U(\Lambda) | \Psi > \, . \label{trasta}
\end{eqnarray}
The Correspondence Principle requires that any field observable $\phi (x)$ which classically transforms as
\begin{eqnarray}
\phi' (x') = S(\Lambda) \phi (x)
\end{eqnarray}
corresponds to an operator ${\hat \phi} (x)$ obeying
\begin{eqnarray}
U(\Lambda) {\hat \phi} (x) U^\dagger (\Lambda) = S^{-1} (\Lambda) {\hat \phi} (\Lambda x) \, . \label{relcond}
\end{eqnarray}
If, instead of a pure state, ${\cal O}$ is in the presence of a density matrix ${\hat \rho}$
\begin{eqnarray}
{\hat \rho} = \sum_i |i> p_i <i| \, ,
\end{eqnarray}
in which he detects the state $|i>$ with probability $p_i$, ${\cal O}'$ will describe the same situation in terms of the transformed density matrix 
\begin{eqnarray}
{\hat \rho}' = U(\Lambda) {\hat \rho} U^\dagger (\Lambda) = \sum_i U (\Lambda) |i> p_i <i|U^\dagger (\Lambda) \, . \label{trasfdens}
\end{eqnarray}
In many applications the Canonical Ensemble is used, defined by the density matrix
\begin{eqnarray}
{\hat \rho}_{C.E.} = \frac {e^{-\beta H}} {Tr e^{-\beta H}} \, . \label{canonical}
\end{eqnarray}
In order to be able to implement the condition
\begin{eqnarray}
Tr {\hat \rho} = 1
\end{eqnarray}
we must imagine that the statistical system under consideration extends over a finite physical volume $V$.

$V$ is the volume of the container and has nothing to do with the volume of the ambient physical space, which, in order to maintain the Lorentz covariance of the theory, will be kept of infinite extension.

\section{Spinless Particles}

We will discuss in this section the case of spin zero particles, which, for simplicity, we assume massless.

The covariantly normalized annihilation operators ${\tilde a} (k)$ are related to those normalized to a Dirac $\delta$-function, $a(k)$, by\cite{huang}
\begin{eqnarray}
{\tilde a} (k) \equiv \sqrt{\omega_{\bf k}} \, a (k) \, .
\end{eqnarray}
where $\omega_{\bf k} \equiv |{\bf k}|$ is the particle energy.

It is easy to check that
\begin{eqnarray}
U (\Lambda) {\tilde a} (k) U^\dagger (\Lambda) = {\tilde a} (\Lambda k) \, . \label{trasfann}
\end{eqnarray}
We are now in the position to discuss the transformation properties of various observables.

A basic quantity is
\begin{eqnarray}
&&E({\bf k}) \equiv  <{\tilde a}^\dagger (k) {\tilde a} (k) > = \nonumber \\
 &&= \omega_{\bf k} Tr ({\hat \rho} \, a(k)^\dagger a (k)) \equiv n({\bf k}) \omega_{\bf k} \, , \label{enedens}
\end{eqnarray}
where
\begin{eqnarray}
d N ({\bf k}) = n({\bf k}) d {\bf k}
\end{eqnarray}
gives the number of particles with momentum within the volume $d {\bf k}$ around ${\bf k}$, in the space volume $V$.

We stress the fact that $E({\bf k})$ and $n({\bf k})$ are extensive quantities, proportional to the volume $V$ of the system.

$E({\bf k})$, defined in eq.(\ref{enedens}) gives the energy distribution among the different modes of the system.

In the ${\cal O}'$ frame, using eqs.(\ref{trasfann}) and (\ref{trasfdens}), we have
\begin{eqnarray}
&&E' ({\bf k}') = Tr ({\hat \rho}' {\tilde a} (k')^\dagger {\tilde a} (k')) = \label{trasfen} \\
&&=Tr (U(\Lambda) \, {\hat \rho} \, U^\dagger (\Lambda) {\tilde a} (k')^\dagger {\tilde a} (k')) = \nonumber \\
&&=Tr ({\hat \rho} \, U^\dagger (\Lambda) {\tilde a} (k')^\dagger {\tilde a} (k') U(\Lambda)) = \nonumber \\
&&= Tr ({\hat \rho} \, {\tilde a} (\Lambda^{-1} k')^\dagger {\tilde a} (\Lambda^{-1} (k')) = E(\Lambda^{-1} {\bf k}') = E ({\bf k}) \, , \nonumber
\end{eqnarray}
where $k' = \Lambda k$.

Recalling the Lorentz invariance of the measure $\frac {d {\bf k}} {\omega_{\bf k}}$, eq.(\ref{trasfen}) shows that $E ({\bf k})$ behaves as a scalar quantity and corresponds to the statement that
\begin{eqnarray}
&&d N ({\bf k}) = n({\bf k}) d {\bf k} = \omega_{\bf k} n({\bf k}) \, \, \frac {d {\bf k}} {\omega_{\bf k}} = \nonumber \\
&&= \omega_{{\bf k}'} n'({\bf k}') \, \, \frac {d {\bf k}'} {\omega_{\bf k}'}  = d N' ({\bf k}') \label{invspinzero}
\end{eqnarray}
is itself a scalar, so that $\int d N$, the total number of particles in the system, is Lorentz invariant\cite{pauli}.

On the basis of eq.(\ref{trasfen}) we can easily determine the transformation properties of any other quantity of interest. 

Consider, as an example, the spectral density $e ({\bf k})$, defined, in terms of $E ({\bf k})$ by
\begin{eqnarray}
&&V e ({\bf k}) d \omega \, d \Omega \equiv n({\bf k}) \omega_{\bf k} d {\bf k} = E ({\bf k}) d {\bf k} = \nonumber \\
&&= E ({\bf k}) \, \omega^2 d \omega \, d \Omega \, ,
\end{eqnarray}
i.e.
\begin{eqnarray}
e ({\bf k}) \equiv \frac {E ({\bf k})} {V} \omega^2 \, . \label{defspec}
\end{eqnarray}
The observer ${\cal O}'$ defines $e' ({\bf k}')$, correspondingly, as
\begin{eqnarray}
e' ({\bf k}') = \frac {E' ({\bf k}')} {V'} \omega'^2 \, ,
\end{eqnarray}
where $V'$ is the physical volume of the container as seen from ${\cal O}'$, which suffers a Lorentz contraction with respect to $V$. If we refer to a situation in which ${\cal O}'$ moves with respect to ${\cal O}$ with velocity ${\bf v}$, we have
\begin{eqnarray*}
&&V = \gamma V' \\
&&\omega = \gamma (1 + {\hat {\bf k}'} \cdot {\bf v}) \, \omega' \\
&&\gamma =\frac {1} {\sqrt{1- v^2}}
\end{eqnarray*}
and, from the Lorentz invariance of $E({\bf k})$, it follows
\begin{eqnarray}
&&e' ({\bf k}') = \frac {E' ({\bf k}')} {V'} \omega'^2 = \frac {1} {\gamma (1 + {\hat {\bf k}'} \cdot {\bf v})^2} \frac {E ({\bf k})}  {V} \omega^2 = \nonumber \\
&&= \frac {1} {\gamma (1 + {\hat {\bf k}'} \cdot {\bf v})^2} e ({\bf k}) = \frac {1} {\gamma (1 + {\hat {\bf k}'} \cdot {\bf v})^2} e (\Lambda^{-1} {\bf k}')  \, .
\end{eqnarray}

\section{Photons}

Real photons also carry a polarization and we will discuss the treatment of this degree of freedom in this section.

We remind that, in the Gupta-Bleuler relativistic formulation\cite{mandl}, photons are described by the four vector potential
\begin{eqnarray}
&&A^\mu(x) = \\
&&= \int \frac {d {\bf k}} {\sqrt{2 (2 \pi)^3} \, \omega_{\bf k}} ({\tilde a}^\mu (k) e^{-ikx} + {{\tilde a}^\mu (k)}^\dagger e^{ikx}) \nonumber
\end{eqnarray}
and the covariantly normalized annihilation operator for a photon of momentum $k$ and polarization $\epsilon (k)$ is given, with an appropriate choice of phases, by
\begin{eqnarray}
{\tilde a}_\epsilon (k) = \epsilon_\mu (k) {\tilde a}^\mu (k) \, .
\end{eqnarray}
In this covariant formalism four kinds of photons exist out of which only two are physical, namely those which correspond to polarization vectors satisfying
\begin{eqnarray}
&&\epsilon_0 (k) = 0 \label{transver1} \\
&&\epsilon_\mu (k) \, k^\mu = \epsilon_i (k) \, k^i = 0 \, . \label{transver2}
\end{eqnarray}
The space components of  the four vector $\epsilon (k)$, i.e. $\epsilon^i (k)$, identify the polarization direction and photons corresponding to $\epsilon$'s not satisfying eqs.(\ref{transver1}) and (\ref{transver2}) are decoupled imposing, on physically realizable states, the condition
\begin{eqnarray}
k_\mu {\tilde a}^\mu (k) |Phys> = 0 \, . \label{physgupta}
\end{eqnarray}
As a consequence of eq.(\ref{relcond}), applied to $A^\mu(x)$, we have
\begin{eqnarray}
U(\Lambda) {\tilde a}^\mu (k) U^\dagger(\Lambda) = ({\Lambda^{-1})^\mu}_\nu {\tilde a}^\nu (\Lambda k) \, . \label{lorcovspin1}
\end{eqnarray}
If, in analogy with eq.(\ref{enedens}), we define
\begin{eqnarray}
H^{\mu \nu} ({\bf k}) \equiv < {{\tilde a}^\mu (k)}^\dagger {\tilde a}^\nu (k) > = \omega_{\bf k} < { a^\mu (k)}^\dagger a^\nu (k) > \, ,
\end{eqnarray}
the Gupta-Bleuler condition, eq.(\ref{physgupta}), implies
\begin{eqnarray}
k_\mu H^{\mu \nu} (k) = k_\nu H^{\mu \nu} (k) = 0 \, . \label{gupta}
\end{eqnarray}
The physical interpretation of $H^{\mu \nu} (k)$ is that
\begin{eqnarray}
&&d N_\epsilon ({\bf k}) = n_\epsilon ({\bf k}) d {\bf k} \equiv \\
&&\equiv \epsilon_\mu (k) \epsilon_\nu (k) < { a^\mu (k)}^\dagger a^\nu (k) > d {\bf k} = \frac {1} {\omega_{\bf k}} \epsilon_\mu (k) \epsilon_\nu (k) H^{\mu \nu} (k) d {\bf k} \nonumber
\end{eqnarray}
gives the number of photons with momentum within the volume element $d {\bf k}$ around ${\bf k}$ {\em and} polarization $\epsilon$, inside the full volume $V$ of the system. Of course the polarization four vector $\epsilon (k)$ has to satisfy the conditions given in eqs.(\ref{transver1}) and (\ref{transver2}).

Lorentz covariance, eq.(\ref{lorcovspin1}), requires, in analogy with eq.(\ref{trasfen}),
\begin{eqnarray}
&&{H^{\mu \nu}}' (k') = Tr({\hat \rho}' {{\tilde a}^\mu (k')}^\dagger {\tilde a}^\nu (k')) = \nonumber \\
&&= < U^\dagger (\Lambda) {{\tilde a}^\mu (k')}^\dagger {\tilde a}^\nu (k') U(\Lambda) > = \nonumber \\
&&={\Lambda^\mu}_\rho {\Lambda^\nu}_\sigma {H^{\rho \sigma}} (k) \, , \label{lorcovspin}
\end{eqnarray}
where, as before, $k' = \Lambda k$.

From eq.(\ref{lorcovspin}) we have
\begin{eqnarray}
&&H'_{\epsilon'} ({\bf k}') \equiv \epsilon_\mu' (k') {H^{\mu \nu}}' (k') \epsilon_\nu' (k') = \nonumber \\
&&=\epsilon_\mu' (k') \epsilon_\nu' (k') {\Lambda^\mu}_\rho {\Lambda^\nu}_\sigma {H^{\rho \sigma}} (k) = \nonumber \\
&&= \epsilon_\rho (k) \epsilon_\sigma (k) {H^{\rho \sigma}} (k) \equiv H_\epsilon ({\bf k}) \, , \label{relpolinv}
\end{eqnarray}
where
\begin{eqnarray}
\epsilon_\rho (k) \equiv \epsilon' (k') _\mu {\Lambda^\mu}_\rho \label{loreps1}
\end{eqnarray}
and eq.(\ref{relpolinv}) would correspond to the intuitive condition
\begin{eqnarray}
d N'_{\epsilon'} ({\bf k}') = d N_\epsilon ({\bf k}) \, , \label{invpolnumb}
\end{eqnarray}
thus extending eq.(\ref{invspinzero}) to the spin-$1$ case. However we still have to face the fact that, in order to be physical, the photon polarization four vector $\epsilon (k)$, as its Lorentz transform $\epsilon' (k')$, must both satisfy eqs.(\ref{transver1}) and (\ref{transver2}), which seems not to be compatible with eq.(\ref{loreps1}).

To see how one can get around this problem, we consider $\epsilon_\rho (k) = \epsilon_\mu' (k') {\Lambda^\mu}_\rho$, for a physical $\epsilon_\mu' (k')$. We have, from eq.(\ref{transver2}),
\begin{eqnarray}
k^\rho \epsilon_\rho (k)  = k^\rho \epsilon_\mu' (k') {\Lambda^\mu}_\rho = \epsilon_\mu' (k') \, {k^\mu}' = 0 \, ,
\end{eqnarray}
which shows that $\epsilon_\rho (k)$ can be decomposed as
\begin{eqnarray}
\epsilon_\rho (k) = {\tilde \epsilon}_\rho (k) + \alpha k_\rho \, , \label{gaugetra}
\end{eqnarray}
where ${\tilde \epsilon}_\rho (k)$ is such that $k \cdot {\tilde \epsilon} (k)=0$ and $\alpha$ can be chosen so that ${\tilde \epsilon}_0 (k) = 0$. In this way ${\tilde \epsilon}_\rho (k)$ has the right form to be interpreted as the polarization of a physical transverse photon and, as a consequence of eq.(\ref{gupta}), can be used in eq.(\ref{relpolinv}) in the place of $\epsilon_\rho (k)$. Therefore eq.(\ref{invpolnumb}) is equivalent to
\begin{eqnarray}
d N'_{\epsilon'} ({\bf k}') = d N_{\tilde \epsilon} ({\bf k}) \, ,
\end{eqnarray}
where $\epsilon' (k')$ and ${\tilde \epsilon} (k)$ both correspond to polarizations of real, transverse photons. Within the framework of the Gupta-Bleuler formulation, eq.(\ref{gaugetra}) is just the statement that the physical polarization ${\tilde \epsilon} (k)$ is the Lorentz transform of $\epsilon' (k')$, up to a residual gauge transformation\cite{mandl}.

We can now proceed as in the spinless case and deduce the transformation properties of any quantity of interest.

For example, the spectral density for a given polarization, $e_\epsilon ({\bf k})$, defined, in analogy with eq.(\ref{defspec}), as
\begin{eqnarray}
&&V e_\epsilon ({\bf k}) d \omega \, d \Omega \equiv n_\epsilon ({\bf k}) \omega_{\bf k} d {\bf k} = H_\epsilon ({\bf k}) d {\bf k} = \nonumber \\
&&= H_\epsilon ({\bf k}) \, \omega^2 d \omega \, d \Omega \, ,
\end{eqnarray}
giving
\begin{eqnarray}
e_\epsilon ({\bf k}) \equiv \frac {H_\epsilon ({\bf k})} {V} \omega^2 \, , \label{defspec1}
\end{eqnarray}
transforms like
\begin{eqnarray}
e_{\epsilon'}' ({\bf k}') = \frac {1} {\gamma (1 + {\hat {\bf k}'} \cdot {\bf v})^2} e_{\tilde \epsilon} (\Lambda^{-1} {\bf k}')  \, .
\end{eqnarray}

\section{Conclusions}

In this paper we presented a systematic method to relate the statistical distributions of quantum systems, as seen from different observers connected by a Lorentz transformation. Although the results are not new\cite{pauli}, we think that our presentation helps in clarifying the hypotheses underlying such issues and also the corresponding formal manipulations.

We applied  the general formalism to the photon case, giving a detailed treatment of the Lorentz transformation of the polarization distribution.

\end{document}